# A Landscape Study of Open Source and Proprietary Tools for Software Bill of Materials (SBOM)

Mehdi Mirakhorli, Derek Garcia, Schuyler Dillon, Kevin Laporte, Matthew Morrison, Henry Lu, Viktoria Koscinski, Christopher Enoch , *Contact: Mehdi23@hawaii.edu, University of Hawaii at Manoa*

*Abstract—*

*Modern software applications heavily rely on diverse third-party components, libraries, and frameworks sourced from various vendors and open source repositories, presenting a complex challenge for securing the software supply chain. To address this complexity, the adoption of a Software Bill of Materials (SBOM) has emerged as a promising solution, offering a centralized repository that inventories all third-party components and dependencies used in an application. Recent supply chain breaches, exemplified by the SolarWinds attack, underscore the urgent need to enhance software security and mitigate vulnerability risks, with SBOMs playing a pivotal role in this endeavor by revealing potential vulnerabilities, outdated components, and unsupported elements.*

*This research paper conducts an extensive empirical analysis to assess the current landscape of open-source and proprietary tools related to SBOM. We investigate emerging use cases in software supply chain security and identify gaps in SBOM technologies. Our analysis encompasses 84 tools, providing a snapshot of the current market and highlighting areas for improvement.*

Modern software applications heavily depend on numerous third-party components, libraries, and frameworks, often sourced from diverse vendors and open source repositories [8]. As a consequence, securing the software supply chain becomes an intricate challenge. To mitigate this complexity, the adoption of a Software Bill of Materials (SBOM) offers a promising solution [12]. An SBOM acts as a centralized repository, providing a comprehensive inventory of all third-party components, packages, and dependencies used in a software application, thereby enhancing software transparency [16], [17], [1].

Recent supply chain breaches, exemplified by the SolarWinds attack [15], [5], underscore the imperative to enhance software security and mitigate vulnerability risks. SBOMs play a pivotal role in this endeavor by revealing potential vulnerabilities, outdated components, and unsupported elements. The concept of SBOM is not entirely new, with a "bill of materials" (BOM) well-established in the manufacturing industry [3]. This information is essential for quickly combating supply chain threats. The importance of this transparency has been further emphasized by President Biden's Executive Order 14028 [13] and the National Cybersecurity Strategy released in March 2023 [14]. With this Executive Order, SBOM is a critical component of software acquisitions, vendor evaluations, risk assessments, and vulnerability management processes. SBOM uniquely identifies all dependencies (software components) that are present in an application, along with information about each component's origin, version, and licensing. There are many methods for uniquely identifying a software's components. In 2009, the International Organization for Standardization (ISO) and the International Electrotechnical Commission (IEC) published the predecessor to SBOMs, Software ID (SWID) Tagging [11]. Along with SWID, there are other software identification methods suggested for use in SBOM such as Package URL (PURL) [4] and Common Platform Enumeration (CPE) [6], [7]. SWID is primarily used for software inventory management and compliance, PURL is used for locating and downloading software packages, and CPE is used for tracking vulnerabilities and understanding risk exposure.

Nevertheless, SBOM usage brings transparency into software production and keeps vendors accountable for their cyber-hygiene. In an effort to stream-line SBOM creation and usage, a variety of tools addressing different SBOM use cases has emerged, from small open source SBOM generators to software composition analysis tools, resulting in a broad and non-standardized range of tooling. In this paper, we have conducted an extensive empirical analysis using mixed-methods: *qualitative analysis*, *instrumentation* and Plugfest (*tool benchmarking*), to assess the current landscape of open-source and proprietary tools related to SBOM. By analyzing these tools, we investigated the current SBOM landscape, characterized the emerging use cases in the area of software supply chain and have identified gaps in SBOM technologies.

More specifically, in this paper we conducted an extensive review of open source and proprietary tools to investigate the emerging use cases in software supply chain security. This included a detailed review, analysis and demonstration of 84 tools to compile a snapshot of the current market and to identify gaps in SBOM technologies. This study demonstrated the challenges ahead to facilitate pervasive and meaningful adoption of SBOM. We executed an open source **Plugfest**, in which five open source SBOM generation tools were selected to participate in the Plugfest event executed by our team to benchmark the most mature OSS SBOM tools and test their interoperability. This study highlighted a number of significant gaps in the standardization of SBOM, key concepts but also interoperability of existing tools.

The **contributions** of our work are four-fold:

- An empirically grounded characterization of the SBOM tooling landscape and emerging use cases.
- An in-depth discussion of interoperability issues, gaps, and limitations in existing SBOM tools but also opportunities for the industry for further standardization of SBOM concepts.
- We organized an extensive SBOM Plugfest, specifically focusing on open-source SBOM tools to assess their interoperability. This evaluation not only revealed the current tools' limitations but also illuminated potential risks that require mitigation before integrating SBOMs into software acquisition and cyber-risk management processes.
- We developed an extensive set of SBOM tool demonstrations. Through our comprehensive study of SBOM-related open-source tools, we execute each one and create step-by-step usage demonstrations. These recorded demos showcase the features, outcomes, as well as the pros and cons of each open-source tool. They are accessible on our YouTube channel (LearnSBOM[1]), which has garnered over 30,000 views from practitioners and 300 subscriptions from SBOM users in the industry. Additionally, this YouTube channel serves as a platform for interactions with developers who use, develop, or innovate using open-source SBOM tools. While the LearnSBOM YouTube channel was a by-product of this study, it has regularly engaged the developer community in discussions about the capabilities of existing tools, and has enabled us to support developers in a faster adoption of open-source SBOM tools. Additionally, several startups developing proprietary tools have requested us to cover their SBOM tools in LearnSBOM YouTube channel.

To support reproducibility of the findings, all of our data, evaluations, source code, pipelines and prototype will be released publicly at github.com.

## 1. METHODOLOGY

The primary objective of this research was to assess and characterize the current landscape of Software Bill of Materials (SBOM) tools through an exploratory study encompassing both open source and proprietary solutions. The research methodology entailed a systematic approach involving the following steps:

- **Defining the Analysis Scope:** The study's initial step involved delineating the scope of our search. We aimed to identify all relevant open-source and proprietary tools associated with SBOM, software composition analysis and tools designed to enhance software supply chain security.
- **Existing Literature and Interviews:** To gain deeper insights into the sentiment surrounding SBOM, we analyzed existing literature and compiled our findings into common patterns. We also held technical discussions with cybersecurity companies featuring open-ended questions to encourage the emergence of relevant topics and insights.
- **Tool Identification:** We conducted an exhaustive search across a variety of sources, including open source repositories, gray literature, web content, forums, relevant blogs, outreach to industry, and participation in U.S. SBOM working groups involving numerous vendors. Our com-prehensive search of resources utilized a set of carefully chosen keywords and phrases associated with SBOM concepts, the software supply chain, software dependency identification, software identification (SWID, CPE, PURL), tracking known vulnerabilities and related standards like CycloneDX [2] and SPDX [10].

---

[1] https://www.youtube.com/@LearnSBOM/about

- **Open Source Tool Identification:** Open source tools were identified through our exhaustive search across various open-source repositories, forums and relevant blogs, and outreach to open source communities to identify open-source software (OSS) tools associated with SBOM concepts.
- **Proprietary Tool Identification:** For proprietary tools, our methodology differed slightly to accommodate their unique characteristics. In the initial stage, we scoured forums, web-based advertisements, other online resources and reached out to SBOM's industrial working groups in the U.S. to locate proprietary tools pertaining to SBOMs. It's noteworthy that certain proprietary tools may offer free demo versions with limited functionality. Consequently, our approach in this section deviated from that of the open-source analysis.

• **Tool Evaluation Criteria:** The identified SBOM tools were subjected to a thorough evaluation process. This involved a thorough examination of their documentation and features, and code review (if open source). To gauge their practical utility, we executed each tool on one or two selected target projects. These projects were chosen strategically to assess each tool's specific capabilities effectively. Through this evaluation process, we considered a range of essential criteria, including core functionality, usability, interoperability with other tools and systems, and customizability to meet specific organizational needs. For proprietary tools offering free demo versions, we rigorously tested the available functionality. This entailed evaluating the tools based on their advertised capabilities and comparing our findings to the claims made by the respective companies. Some vendors provided us a confidential recorded demo. For tools that did not offer demos, we relied on their publicly available documentation and sought out customer feedback from sources such as product reviews and online forums to corroborate our findings.

• **SBOM Plugfest and Benchmarking:** For the subset of tools that were accessible to us, we executed them on sample software projects. This hands-on approach allowed us to benchmark and compare the performance of these tools, providing valuable insights into their practical utility and effectiveness.
• **Characterization of SBOM Tooling Landscape:** In this stage, we consolidated and analyzed the information gathered from our interviews, tool analysis, and plugfest to understand common similarities and summarize gaps in current open source and proprietary tools.

## 2. Landscape of SBOM Tooling

Through our research we preformed an extensive search of open source projects from independent developers and companies that utilized SBOMs in any form. In total, we analyzed 84 tools that varied in size and function. This section summarizes the results of our empirical analysis of SBOM tools and emerging use cases.

### 2.1. Emerging Use-Cases

We examined the emerging use cases for SBOM tools to gain insights into how these tools can be used to address new and evolving challenges in software supply chain security. We found all tools could be grouped into one or more of the following use cases:

TABLE 1: SBOM Tool Use Cases

| Use Case | Tool Count | Percent of Total |
|---|---|---|
| SBOM Generation | 54 | 64% |
| SBOM Consumption | 31 | 37% |
| SBOM Interoperability | 15 | 18% |
| SBOM Quality Assurance | 14 | 17% |
| SBOM Services | 13 | 15% |

**SBOM Generation:** This use case focuses on generating some form of SBOM. All tools support either one or both SPDX and CycloneDX SBOM schemas, with a select few also using their own SBOM schema developed by the tool maintainers. The majority of this category, 54 total tools, generate SBOMs from project sources. However, 8 tools can be sub-classified as **SBOM Interfaces**: tools that produce SBOMs not from sources, rather user input. These tools provide an interface or framework to create SBOMs and are often used by many traditional SBOM generation tools.

**SBOM Consumption:** This use case focuses on the utilization of SBOM as a resource for supply chain security after generation. This use case includes:

- **Vulnerability Management via Vulnerability Exploitability eXchange (VEX):** This use case focuses on consuming SBOMs to identify possible vulnerabilities in a piece of software to make the developer and client more security-aware. These tools consume SBOMs and query databases like the National Vulnerability Database (NVD) to find vulnerable components and return information about them.
- **Detection of Dependency Misuses:** Tools supporting this SBOM use case focus on detecting dependency misuses that would result in inaccurate SBOMs. Examples of dependency misuses are unpinned dependencies (allowing a mutable version or range of versions), dynamically loading dependencies or, in case of using package management, failing to specify a package's source repository (public vs private repositories), using non-unique version numbers for packages or allowing package managers to fall back to public repositories when internal repositories are unavailable.
- **License Compliance:** This use case focuses on consuming SBOMs and finding licensing issues associated with the software components used in an application.

**SBOM Interoperability:** Software vendors will often need to aggregate multiple SBOMs produced by several software suppliers using different standard formats, into a single SBOM. This creates a problem with interoperability, since SBOMs do not always have a one-to-one mapping between schemas. These use cases focus on analyzing, manipulating, and consuming SBOM artifacts to increase the interoperability of SBOMs generated by different vendors. This category can be further subdivided into *Converting*, *Merging*, and *Comparing* tools.

- **Converting** converts existing SBOMs between different SBOM schemas and file types, accounting for different data fields and the content of the various schemas.
- **Merging** consumes multiple SBOMs to combine their data into a single SBOM.
- **Comparing** compares two SBOMs and outputs the difference to the command line or to a separate SBOM.

**SBOM Quality Assurance:** Tools implementing this use case focused on consuming SBOMs to assess their validity and quality. This use case includes:

- **Schema Checking**, confirming that the SBOM conforms to its respective schema.
- **SBOM Content Validation**, verifying that an SBOM contains the following information: generation information, the software's version/SHA, IDs (PURLs, SWID) of packages/dependencies, versions/SHAs of packages/dependencies, and licensing information of all components.
- **SBOM Scoring**, compiling metric findings in an empirical score.

**SBOM Services:** As SBOMs continue to mature, there is a growing need to SBOM services to facilitate the exchange of SBOM artifacts. This use case focuses on tools that provide the needed infrastructure to support their adoption.

- **SBOM Storage and Management:** Organizations require a centralized location for storing SBOM data, which can either be managed internally or publicly shared. This use case focuses on specialized databases designed exclusively for SBOM management that can utilize SBOMs for vulnerability, dependency, and license management.
- **SBOM Sharing:** The systematic sharing of SBOMs across the software supply chain and vendors is critical for SBOM adoption. Existing open source tools such as DBOM provide open attestation sharing infrastructure for sharing this information between supply chain partners.
- **Incident Response:** This use case automates incident management, where software producers can rapidly notify all vendors about vulnerabilities, and similarly, vendors quickly can find any vulnerable components in their application stack. In particular, existing use cases are around automating communication of Vulnerability Exploitability eXchange (VEXs), where suppliers can share vulnerability information with product recipients and disclose whether a product or products are affected by a known vulnerability.
- **SBOM Signing:** Signing and Notary tools generate an RSA public/private key pair for SBOM signing. The notary has been limited to verifying the signature. These tools do not support identifying discrepancies or omissions in the SBOMs.

> **Classification of Use Cases**
> SBOM use cases are emerging; currently, there is no clearly defined classification of tools and use cases. Without official classifications of tools, it can be challenging to select the right tool for the job. This places the burden of research on individuals trying to adopt SBOM usage. In our online Appendix, we provide the first extensive classification of existing open source SBOM tools.

## 2.2. SBOM Generation

SBOM generation is the essential and most widely supported use case, with over half of the tools ( 54 total ) supporting some form of SBOM generation. Because of this, we placed a large focus on this area of SBOM tooling, conducting an extensive analysis to expose both areas that were well-developed, and those that needed additional support. We demonstrate SBOM generation capabilities and limitations using our Plugfest experiment in section.

### 2.2.1. Supported Inputs and Development Environments

In alignment with the categorization outlined in the U.S. Government's guideline on Types of Software Bill of Material (SBOM) Documents [9], our research categorized the various inputs and development environments supported by SBOM tools into five distinct definitions: **Design**, **Source**, **Build**, **Analyzed**, **Deployed**, and **Runtime**. It's important to note that SBOM tools often leverage one or more of these avenues to generate comprehensive SBOMs.

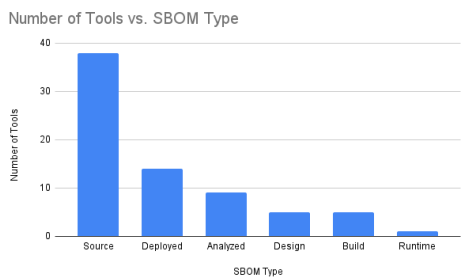

FIGURE 1: Supported SBOM Types for Open Source Tools

**Source SBOMs** are ". . . created directly from the development environment, source files, and included dependencies used to build the product" [9]. We found that 38 tools ( 63% ) support source SBOM generation and we further divided this category into two generation subcategories: via **manifest files** and via **source code**.

- **From Manifest Files: 33 tools.** Manifest files are files containing lists of dependency information, such as pom.xml for Maven or package.json for NPM. They also include "lock" files, which are compressed versions of manifest files often created and modified by the package manager rather than a user. We classified manifest files as source files since without them, projects that utilize build or package management systems that require these fields would not be able to be run. We also made this distinction due to their heavy use by SBOM generation tools. 73%[2] of source SBOM generators require manifest files to function.
- **From Source Code: 10 tools.** Four tools tested explicitly use source code analysis to find dependency information. This process involves traditional software composition analysis techniques and specific elements, such as dependency import statements. However the scope is often limited and manifest files are frequently required to gain any transient dependency information.

**Deployed SBOMs** ". . . contain an inventory of software that is present on a system where software is deployed" [9]. We found that 14 tools ( 26% ) that support deployed SBOM generation. While closely related, the distinction between source and deploy generation is determined by how the tool finds package information. Source SBOM generators use manifest files verbatim, whereas deployed SBOM generators will actively use package managers, either using language package managers such as pip and Maven or distribution package managers such as apt-get and yum, to get package information. Manifest files are not guaranteed to match what is installed on a given system, however by utilizing a package manager, deployed SBOMs capture a snapshot of what is actually installed in a given environment[3].

---

[2] Six of the 33 tools use manifest files, but can use source code

[3] Tools will frequently create an 'adhoc' or actual manifest file by using a package manager to list out the installed dependencies, redirecting it to a file, then using that file as input. We classified these tools as Deployed SBOM generators despite using a manifest file as input because the deployed environment is actively captured during generation time

> **Package Manager Dependence**
>
> SBOM generation tools are exceedingly dependent on package management systems. In isolated environments, such as a virtual Python environment, where only the packages required for the software are installed, SBOM generators can produce robust and accurate SBOMs. In a non-isolated environment where packages are installed globally, SBOM generators will often include any package installed, regardless if the target software uses that package. The resulting SBOMs contain information about the software, but include excessive data about unrelated packages. In the case where no package management system is used, SBOM generators produce SBOMs with little to no data about the target software.
>
> When using manifest files, there is little to no cross-referencing to the source code. Many tools will accept the manifest file verbatim, regardless if the listed dependency is used in the software or not.

**Analysis SBOMs** are created "…through analysis of artifacts (e.g., executables, packages, containers, and virtual machine images) after its build" [9]. We found 9 that tools ( 17% ) that support analysis SBOM generation. These tools are uniquely beneficial to software consumers, who can generate SBOMs without needing direct access to the source code.

**Design SBOMs** are "SBOM of intended, planned software project or product with included components (some of which may not yet exist) for a new software artifact" [9]. We found 5 tools ( 9% ) that could produce Design SBOMs. We classified tools that consumed user input or the definition of an SBOM, such as an SBOM model library, as Design SBOM generators. Many SBOM generators utilize these libraries because they provide a framework for working with SBOMs. However, at this stage no input has been defined. Therefore, these libraries can be used to produce SBOMs for software that doesn't exist using information provided by the user.

**Build SBOMs** are "generated as part of the process of building the software to create a releasable artifact (e.g., executable or package) from data such as source files, dependencies, built components, build process ephemeral data, and other SBOMs" [9]. We found 5 tools ( 9% ) that support build SBOM generation. These tools often take the form of plugins for build systems, such as Maven or Gradle, and run as a step during the compilation process. Build SBOMs provide a clearer insight into what is actually included inside the final binary than a source or analyzed SBOM may be able to provide. The plugin nature of the tools allows for easier integration into CI/CD pipelines and generate an updated SBOM upon new releases. However, especially for proprietary software, the generation responsibility is largely with the maintainer or developer as consumers may not have access to the source code to generate build SBOMs themselves.

**Runtime SBOMs** are "…instrumenting the system running the software, to capture only components present in the system, as well as external call-outs or dynamically loaded components" [9]. This category has the weakest tool support, we found one tool that supported runtime SBOM generation. The benefit of runtime SBOMs is that it contains the dependencies the program uses during runtime, creating a "dynamic" SBOM that represents software in real-time.

### 2.2.2. Supported Ecosystems

Of the 49 OSS non-design SBOM generators we examined, we found a total of 15 languages and 35 package management or build systems were supported[4] (See Fig. 2). Additionally, we found eight and four tools that support a Linux or Windows package manager, respectively. Four tools were capable of supporting image binaries. The breakdown of specific supported systems can be seen in our replication package (to be released).

### 2.2.3. Supported Schemas

CycloneDX, developed by OWASP and SPDX, developed by the Linux Foundation, are the prevailing SBOM formats and the majority of generation tools produce at least one of these schemas. Other SBOMs include more localized and less adopted schemas, often only used by the tool itself. These third-party SBOMs range from well formatted specifications to well organized lists of dependencies. SBOM standards are continuing to evolve and the closest universal standard for SBOMs are the seven minimum elements of an SBOM as defined by the NTIA [12], which include supplier names, component names, component versions, unique identifiers, dependency relationships, authors, and timestamps.

Of the two most common SBOM standards, CycloneDX is the more prevalent schema and supported by 40 out of 54 tools we reviewed. Additionally, third party generation tools not developed by CycloneDX or SPDX are more likely to support CycloneDX over SPDX. Given a third party tool, there is an 86% chance it supports CyloneDX and 55% change it supports SPDX.

---

[4]Binaries are included as a packaged system.

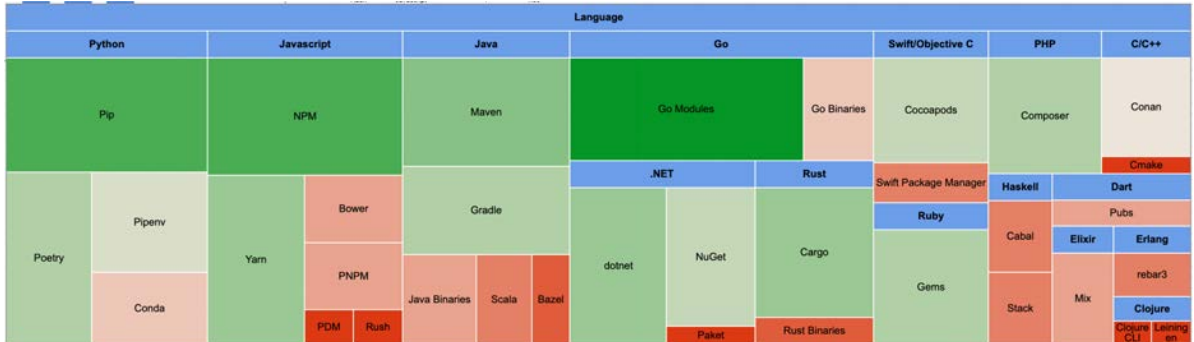

FIGURE 2: Supported Ecosystems

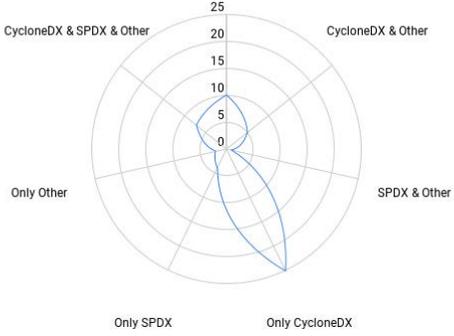

FIGURE 3: SBOM Schema Support

> **Disproportionate Technology Support**
> Language support varied across different tools, however the majority of tools have a highly niche focus. Out of the 54 SBOM generators, 40 ( 74% ) produce only one SBOM type, 31 ( 51% ) produce one schema, and 40 ( 74% ) support one language. Conversely, only five tools ( 9% ) supported more than 10 languages, four of which are part of the seven tools that can generate more than one SBOM type and schema. This disparity results in developers either depending on a small collection of tools for all of their projects or utilizing a different niche tool for each project.

## 2.3. SBOM Consumption and Utilization

SBOM consumption and utilization tools take SBOM artifacts as input and utilize them for other purposes. We found a total of 31 tools that utilize SBOMs in this manner.

*2.3.1. License Management*

*2.3.2. Vulnerability Management and VEX*  Vulnerability management is large use case for SBOM and tools that use SBOM. Vulnerability scanners help automate the detection and remediation of software vulnerabilities. These scanners check software components against vulnerability databases and report the findings. We found nine tools that directly consume SBOMs to produce vulnerability reports However, we did find an additional seven tools that strictly used manifest files as input. We have excluded these tools from the total tool count from this section as manifest files are not SBOMs, but are closely related and their inclusion demonstrates the value of a SBOM-like document for vulnerability management.

Among the vulnerability scanners, 12 tools utilized VEX to some capacity. Six can be classified as **VEX generators** and produced VEX-like vulnerability reports in CycloneDX, OpenVEX, or CSAF VEX standards. Four tools provided a **VEX interface**, allowing users to modify and update VEX statements. The remaining two tools were the only cases of **VEX utilization** we found. These tools could generate vulnerability reports from sources and SBOMs, then update the final report utilizing the information reported in VEX.

*2.3.3. Miscellaneous Use Cases*  Seven SBOM tools are quality of life utilities that make working with SBOMs easier. These tools include features such as visualization and access management.

## 2.4. SBOM Interoperability

SBOM interoperability presents a complex roadblock to widespread SBOM adoption. CycloneDX and SPDX schemas support specific use cases, however do not have a direct one to one mapping, preventing collaboration between vendors using different schemas. We found 15 tools ( 18% ) that addressed this issue by providing means to convert, merge, and compare CycloneDX, SPDX, and third-party SBOMs. Many of these tools supported a third-party SBOM, ranging from adhoc to a well defined schema, that was utilized internally and acted as a reference model for converting to other schemas. A common workflow would include generating a third-party SBOM and providing utilities to convert it to into CycloneDX or SPDX.

> **Limited Standardization**
> Although there are minimum required fields for an SBOM, there is not a universally agreed-upon format. This creates many interoperability issues, making certain SBOMs incompatible with certain tools or functionalities. Even in case of mainstream formats, SPDX does not map directly to CycloneDX and vice versa. A growing consequence are intermediary SBOMs, or SBOMs that are purely designed as an interface between CycloneDX and SPDX. Without clear guidelines from these schemas, intermediary SBOMs can interpret different fields as equivalents, creating inconsistent converted SBOMs.

## 2.5. SBOM Quality Assurance

SBOM quality assurance tools attempt to provided metrics and scores for SBOMs to judge how trustworthy they are. High quality SBOMs are essential if they are be used as a trusted resource for license or vulnerability management. We found 14 tools ( 17% ) that provided methods to verify SBOM quality. The majority of tools, nine ( 64% ) of this category only provide schema validation, whereas the remain five tools provide actual metrics. Specific metrics vary slightly between tools, however in addition to schema validation, common metrics include:

- **NTIA Compliance:** Ensure that the minimum elements of a SBOM as defined by the NTIA [12] are present.
- **License Validation:** Ensure that licenses are present and if they are, confirm they are valid and non-conflicting. The SPDX License List is usually the source of the license information as they are accepted and used by both CycloneDX and SPDX standards.
- **UID Validation:** Ensure that UIDs, such as CPEs and PURLs, are present and correctly formatted.
- **Hash Presence:** Ensure that component hash information is present.
- **Dependency and Relationship Information:** Ensure that the SBOM contains a dependency tree that shows how its components relate to each other.

Some tools also provide custom weighting systems for each metric.

> **Lack of Standardized and In-Depth SBOM Quality Metrics**
> Although there are common community-developed metrics to verify the quality of a tool and the generated SBOM. The effectiveness of a tool cannot be properly assessed without consistent metrics and accurate benchmark tests. In the short term, tool developers and consumers need a common framework to test and validate the SBOM's quality properly. Additionally, metrics are limited to the scope of the SBOM document. SBOM quality assurance is often done without the context of the source code, resulting in insightful cross referencing metrics such as hash validation being impossible.

## 2.6. SBOM Services

SBOM services provide key infrastructure to facilitate SBOM adoption. The impact of high grade SBOMs is limited without the means to distribute them securely or incorporate them into the CI/CD process. We found 13 tools ( 15% ) that provided a service for SBOM. These include repository servers, CI/CD plugins, and signing and attestation tooling.

# 3. SBOM Plugfest: A Benchmark Study of Open Source SBOM Tools

This section reports a benchmark study of existing tools for generating SBOMs. This analysis was conducted using Open Source SBOM generation tools to showcase the differences between SBOMs generated for a control Java project using Maven.

## 3.1. Study Setup and Methodology

*3.1.1. Tool Selection*   In total, five tools were chosen for Plugfest. We chose tools that ranked highest in the amount of languages and ecosystems they support as they cover the largest amount of use cases and more likely to be used by developers.

TABLE 2: Plugfest Tools

| Tool | Vendor | Supported Lang & Ecosys |
|---|---|---|
| CycloneDX Generator | CycloneDX | 14 |
| Syft | Anchore | 14 |
| Trivy | Aqua Security | 12 |
| OSS Review Toolkit | OSS Review Toolkit | 10 |
| SPDX SBOM Generator | Independent | 9 |

*3.1.2. SBOM Type* Each SBOM generated will be a **source SBOM**. Since the most common tool type and use case is source SBOM generation, we focused on this area specifically to test common scenarios discussed in the next section,

*3.1.3. Benchmark Project and Test Cases* A simple example Java project was created using the Maven build system and native Java to demonstrate how each tool performs in the following scenarios:

1) **"As Intended"** This describes a normal project where dependencies are imported and used as intended. Dependencies are present in the manifest file, imported into the source code, and actively used.
2) **"Dead Imports"** This describes a scenario where due to refactoring or updating code, dependencies are imported but never utilized. Dependencies are present in the manifest file, imported into the source code, however **not** actively used.
3) **"Manifest Only"** This describes a more extreme refactoring scenario where code or source files removed, however the manifest still contains the dependency. Dependencies are present in the manifest file, however **not** imported into the source code and **not** actively used.

Figure 4 displays the compile time dependency graph of the control Java project used. **Direct dependencies** are directly referenced in the source code and listed in the manifest file, whereas all **transient dependencies** are required to use the direct dependencies, but are not directly referenced in the source code and do not appear in the manifest file.

*3.1.4. Study Execution and Methodology* To test these tools in an isolated and consistent environment, we developed a standardized Plugfest Docker image that was subsequently used by each tool image. Scenario one was run twice, the first of which Maven was preinstalled in the Plugfest environment and the latter without Maven. Scenario two was run once using without Maven installed in the Plugfest environment. Each tool image installed itself and ran against same Java control project for the given case. Although all tools supported CycloneDX or SBOMs, any tool that generated a third-party SBOM as default was kept in their original formats a tool used, even if there was generation or conversion support for CycloneDX or SPDX to prevent any data loss with conversion. When analyzing the resulting SBOMs, we searched for the minimum elements of an SBOM as defined by the NTIA. The element fields are marked as present if they are **consistently extractable** from the SBOM.

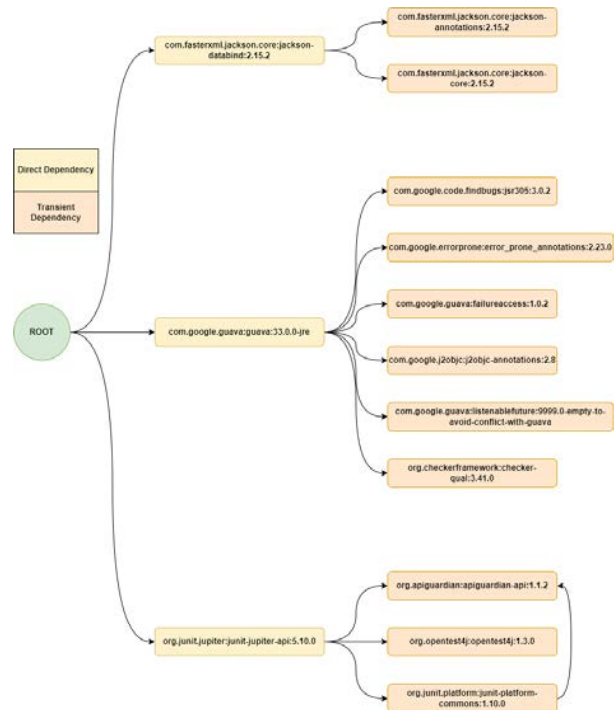

FIGURE 4: Control Java Project Dependency Graph

3.2. Plugfest Results and Findings

The Plugfest results can be found in Figure **??**. Green indicates the value was present, yellow indicates the value was not found, and red indicates the value was present but incorrect. After the reviewing the results, we found the following:

**Finding 1: Source Code has no Impact on SBOM Output.** Each test case was designed to test edge cases for software composition analysis to determine the impact the source code had on the final SBOM. However for each scenario and excluding trivial differences such as timestamps, **each tool produced** the identical SBOMs for each use case. Component information remained the same, regardless of any change in the source code. This raises a major concern for SBOM generation as SBOMs are intended to be a representation of the software and this finding demonstrates the risk of accepting a manifest without cross-referencing source code.

**Finding 2: There is a large disparity between the usage of PURLs and CPE.** PURLs for components were far more common the CPEs. Syft was the only tool that provided CPE and PURL information for components, whereas CycloneDX Generator, Trivy, and OSS Review Toolkit only provided PURL information, and SPDX SBOM Generator provided neither PURLs or CPEs.

**Finding 3: Maven and manifest files are required for dependency relationships.** The presence of Maven and the pom.xml manifest file greatly impacted the dependency capabilities of the tools. In scenario one when Maven was installed in the Plugfest environment, CycloneDX Generator, Trivy, and OSS Review Toolkit successfully reported direct and transient dependencies. This was accomplished by either by one of two methods: directly using Maven or querying the Maven central repository to obtain package information. CycloneDX Generator utilized the former, which is why in scenario one when Maven was not installed, only information stored in the pom.xml was parsed and resulted in the lost transient dependency information. SPDX SBOM Generation also utilized Maven, however its generation capacity was far weaker and produced 12 false positives not found by any other tool. It also could not run without Maven installed. Trivy and OSS Review Toolkit utilize the latter, demonstrated with consistent relationship information regardless if Maven was installed or not. Syft did not use either of these methods and depended solely on the pom.xml file, which contained only direct dependencies and reflected in the resulting SBOM.

The loss of dependency relationships is best represented in scenario 2. The greater loss of information is discussed in Finding 4, however **all direct and transient relationship information is lost in scenario 2.**[5] Without Maven or pom.xml manifest, determining dependency relationships becomes increasingly more difficult.

**Finding 4: The Removal of Build Systems Reduces the Amount of Data in the SBOM.** Initially with Maven and the manifest file available, the total measured missing data was 35%. However when those build resources were removed in scenario 2, the missing data from the SBOMs produced in scenario two by CycloneDX Generator, Syft, Trivy, and OSS Review Toolkit grew to 65%.

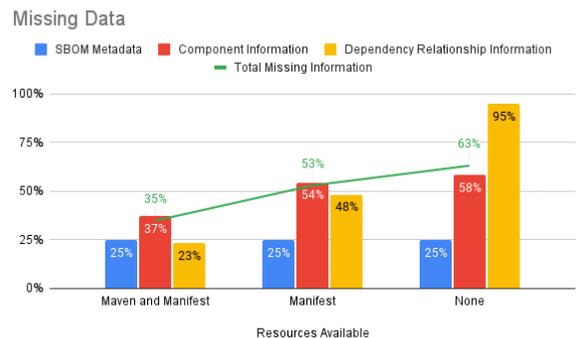

FIGURE 5: Missing Data with Different Resources[6]

Figure **??** demonstrates the trend of data loss as build systems are removed. SBOM metadata is predictably stagnate as that is determined by the tool itself. Component information levels off since even without Maven, the jar files provide some additional information. ClclondeDX Generator and Syft drastically increase reported component information from scenario 1 with Maven to scenario 2 since the data source changes from the manifest file to the dependency jar themselves.

Most notable is the exponential growth of the loss of dependency relationship information. When compiling from sources without a build system, nearly all relationship information is lost. This is critical to dependency and vulnerability management as without relationship information, removing specific problem dependencies becomes increasing costly and complex.

---

[5]Syft-style SBOMs use a "contains" relationship keyword to denote dependencies ( i.e. source "contains" dependency) and does report that the source folder contains all the jar dependencies. However, Syft refers to directory structures rather than applications and while the source **directory** does "contain" all the needed dependencies to run, it's unclear what the source **application** and its transient dependencies depend on. As a result, we reported that Syft found direct dependency relationships in figure **??** since the both source directory and application "contain" those dependencies. We did not report transient dependency relationships even if transient dependency was defined because although not incorrect, "contains" is too ambiguous to determine transitive dependency relationships.

### 3.3. Overall Analysis

This benchmark analysis revealed numerous gaps in existing tools that support generation of SBOMs. In particular, we observed trends of data loss and cases package manager dependence. The SBOMs generated by these tools all contained similar information about the control project, but varied in the degree of thoroughness.

**Lack of Ground Truth:** A significant challenge is lack of large scale and extensive ground truths to compare SBOMs against. The quality, accuracy, and completeness of an SBOM generator tools at scale cannot be assured without the ability to validate them with complex ground-truth data that cover various forms of dependency use and misuses.

**Accuracy in Case of Dependency Malpractices:** Current tools rely heavily on secondary systems such as package managers or manifest files to generate SBOMs. Secondary systems may report information not reflected in the source code. There are cases where developers may use bad dependency practices (e.g., hard-coding or dynamically loading dependencies), creating an inaccurate SBOM. This may spark unnecessary remediation strategies or provide a false sense of security.

**Inconsistencies:** The usage of Open Source SBOM generator tools created many inconsistencies for the SBOMs generated. Even tools that are compatible with multiple languages may report different SBOM fields per language, such as providing suppliers for one language but not another language.

**Lack of Standardization for SBOM Ingredients:** A lack of standardization on what forms of dependencies must be included in SBOMs creates ambiguity and confusion for SBOM generation and usage. Some tools report imported and unused dependencies, some report transient dependencies and some tools included a multitude of system, runtime, security, and collections packages in their SBOM. This raised questions as to what components should be included or excluded within an SBOM.

**Limited Support for Variety of Package Management Systems:** The availability of tools per language and methods of generation per language further complicates SBOM generation. Languages with multiple package managers had some limitations with Open Source tools where some tools only provided functionality for one package manager. For some languages, this may hinder the usage of SBOMs, as their specified package manager could have limited accessibility to SBOM generation tools.

**Software Identification Challenges:** Current minimum SBOM requirements leave room for interpretation of data fields. Multiple types of information can satisfy the same requirement, such as SWID and CPE both fitting the software ID category. It is up to the tool or SBOM schema developer's discretion on what type to use. There is also a lack of standards for any additional fields beyond the minimum requirements. This can cause SBOMs of different schemas to be incompatible with each other and make it difficult to work with multiple schemas at once.

## 4. CONCLUSION

In this study, we collected data on 84 SBOM-related tools to assess the current market. We identified common trends in these tools to create a set of classifications based on use cases. The wide range of SBOM tools is an encouraging sign that the software development community is actively embracing the SBOM concept and software transparency. However, sifting through this growing field of applications can quickly overwhelm developers and product owners who are eager to adopt SBOMs. This is especially difficult as each application can accomplish one or more of the diverse tasks related to SBOM usage. Additionally, the needs of SBOM users are unique and diverse. This makes it necessary to identify the use cases of SBOM tools and their possible gaps and shortcomings. Remediating the gaps in the current SBOM market is essential to increasing SBOM adoption and ease of use. Most OSS tools require manual generation of SBOMs. Maintaining an updated SBOM after code changes requires the generator to be rerun against that changed code. Some tools (mainly proprietary) offer continuous integration through repositories like GitHub, GitLab, and BitBucket. These tools can utilize features such as GitHub Actions to automate SBOM generation or vulnerability scans. Other tools create or update SBOM information. These tools automatically run when changes are saved to a project or pushed to a repository, ensuring SBOMs and reports are always up to date. Software Supply Chain Security is a growing topic. This paper highlights a snapshots of SBOM tooling landscape a year after President Biden's Executive Order on Cybersecurity. As of today, further research and development is needed to successfully integrate SBOM tooling into software acquisition processes. This paper aims to highlight limitations of current tools and serve as an avenue for further dialog among practitioners and researchers.

The growing open source SBOM tool market suggests increased SBOM interest and adoption. However, the current SBOM market still has a few gaps that must be addressed for SBOM adoption to be completely streamlined.